\documentclass[aps,prd,amsmath,amsfonts,amssymb,eqsecnum,nofootinbib,floatfix,secnumarabic,preprintnumbers,superscriptaddress,nobalancelastpage,onecolumn,notitlepage,biblatex,12pts]{article}
\pdfoutput=1 % if your are submitting a pdflatex (i.e. if you have
\usepackage[T1]{fontenc} % if needed

\usepackage[]{caption}
\usepackage{makeidx}
\usepackage{multirow}
\usepackage{multicol}
\usepackage[dvipsnames,svgnames,table]{xcolor}
\usepackage{graphics}
\usepackage{epstopdf}
\usepackage{ulem}
\usepackage{hyperref}
\usepackage{amsmath}
\usepackage[compat=1.0.0]{tikz-feynman}
\usepackage{amssymb}
\usepackage{float}
\usepackage{fancyhdr} 
\usepackage{enumitem}
\usepackage{tcolorbox}
\usepackage{lipsum}
\usepackage{bold-extra}
\usepackage{wrapfig}
\usepackage{caption}
\usepackage[numbers]{natbib}
\allowdisplaybreaks
\usepackage{upgreek}
\usepackage{bm}					% Package for bold math symbols
\usepackage{booktabs} % To thicken table lines

\newcommand{\be}{\begin{eqnarray}}
\newcommand{\ee}{\end{eqnarray}}
\newcommand{\bea}{\begin{eqnarray}}
\newcommand{\eea}{\end{eqnarray}}

\newcommand{\lt}{\left}
\newcommand{\rt}{\right}

\newcommand{\ba}{\begin{array}}
\newcommand{\ea}{\end{array}}
\newcommand{\bd}{\begin{displaymath}}
\newcommand{\ed}{\end{displaymath}}
\newcommand{\beq}{\begin{equation}}
\newcommand{\eeq}{\end{equation}}
\def\b{\beta}

\def\q2 {q^2}

\def\bt{\begin{table}}
\def\et{\end{table}}

\definecolor{shilamagenta}{rgb}{0.8, 0.0, 0.8}

\definecolor{shilagreen}{rgb}{0.0, 0.5, 0.0}
\definecolor{shilacyan}{rgb}{0.0, 0.58, 0.71}
\definecolor{midnightblue}{rgb}{0.1, 0.1, 0.44}

\usepackage{hyperref}

\hypersetup{citecolor=blue,        % color of links to bibliography
    linkcolor=red,  % color of internal links (change box color with linkbordercolor)
    filecolor=magenta,      % color of file links
    urlcolor=shilacyan  
}

\usepackage{cleveref}
\usepackage{slashed}%for slashed term in Dirac equation
\begin{document}

\title{\textcolor{black}{\textbf{The N2HDM, Entropy Production and Stochastic Gravitational Waves}}}

\author{Arnab~Chaudhuri%\thanks{corresponding author}\ 
$^{a}$\footnote{{\bf e-mail}: \href{arnab.chaudhuri@nao.ac.jp}{arnab.chaudhuri@nao.ac.jp} }
and
Kazunori Kohri$^{a,b,c,d}$\footnote{{\bf e-mail}: \href{mailto:kazunori.kohri@gmail.com}{kazunori.kohri@gmail.com}}
\\
$^a$ \small{Division of Science, National Astronomical Observatory of Japan,} 
\\
\small{Mitaka, Tokyo 181-8588, Japan.}\\
$^b$ \small{The Graduate University for Advanced Studies (SOKENDAI), Mitaka, Tokyo 181-8588, Japan.}
\\
$^c$ \small{Theory Center, IPNS, and QUP (WPI), KEK, 1-1 Oho, Tsukuba, Ibaraki 305-0801, Japan}
\\
$^d$ \small{Kavli IPMU (WPI), UTIAS, The University of Tokyo, Kashiwa, Chiba 277-8583, Japan}
}

\date{\today}
\maketitle

\begin{abstract}
This study undertakes a reconsideration of the potential for a first-order electroweak phase transition, focusing on the next-to-minimal two Higgs doublet model (N2HDM). Our exploration spans diverse parameter spaces associated with the phase transition, with a particular emphasis on examining the generation of stochastic Gravitational Waves (GW) resulting from this transition. The obtained results are meticulously compared against data from prominent gravitational wave observatories, and the possibility of their detection in the future GW observations have been established. In passing by we analyse the strength of the phase transition through the production of entropy during the electroweak phase transition.\\
\\
\textbf{Keywords:} N2HDM, Entropy production, Gravitational Waves, Electroweak Phase transition.
\end{abstract}

%-----------------------------------
%.  NEW SECTION
%----------------------------------
\section{Introduction}
\label{sec:intro}
While the Standard Model (SM) of particle physics has achieved remarkable success in elucidating three of the four fundamental forces—namely, electromagnetic, weak, and strong forces—it remains incomplete in its quest to comprehensively describe all known elementary particles and the entirety of fundamental interactions. Despite its popularity and widespread acceptance, the SM grapples with certain unresolved phenomena. Notably, it fails to provide a complete explanation for baryon asymmetry, lacks a cohesive incorporation of the gravitational theory outlined in general relativity, and does not address the universe's accelerating expansion, a phenomenon possibly governed by dark energy. Furthermore, the SM lacks a suitable candidate for dark matter, leaving crucial aspects of our understanding of the cosmos unaccounted for.

The origin of the observed asymmetry between matter and antimatter, known as baryogenesis, adheres to Sakharov's principle \cite{Sakharov:1967dj}, encompassing three essential conditions: \textbf{i) }Non-conservation of Baryon numbers, \textbf{ii) }Breaking of C and CP invariance, and \textbf{iii) }Deviation from thermal equilibrium. To elucidate the baryon asymmetry in the universe through baryogenesis, a pivotal prerequisite is a strong first-order electroweak phase transition (EWPT) during the early universe. The cosmic EWPT transpired as the primeval universe cooled sufficiently, causing the Higgs field's potential to settle at a non-zero minimum. Consequently, the symmetry of the theory $SU(3)_C \times$ $SU(2)_L\times U(1)_Y$ broke into $U(1)_{\rm em}$. At the onset of this first-order EWPT, bubbles of the broken phase emerged, generating baryon-antibaryon asymmetry outside the bubble walls.

However, with the subsequent discovery of the SM Higgs boson \cite{Higgs}, it became apparent that the EWPT within the SM framework is a smooth crossover phase transition (see \cite{Chaudhuri:2017icn, shapas}). Consequently, for a successful electroweak baryogenesis (EWBG), a first-order phase transition theory beyond the SM is imperative. Studies have delved into strong first-order phase transitions in $Z_1$, $Z_2$ and $Z_3$ extensions of the SM \cite{Chaudhuri:2022sis, Morrissey:2012db,z31,z32,z33} to explore viable alternatives. 

In essence, the strength of the EWPT hinges on the intricate interplay between the scalar potential's behavior at high and low temperatures. The critical temperature, determined through computation, serves as a pivotal parameter, revealing shifts in the global minimum within the scalar potential concerning the temperature (T) of the Universe. To gain a precise understanding of the electroweak phase transition (EWPT), it becomes imperative to delve into the dynamics of bubble nucleation. This process, essential for the progression of the phase transition, involves the formation of bubbles \cite{Mazumdar:2018dfl}. Particularly during the first-order EWPT, the dynamics of bubble nucleation can give rise to stochastic gravitational waves (GWs) in the early Universe \cite{Apreda:2001us, Grojean:2004xa, Weir:2017wfa, Ellis:2018mja, Alanne:2019bsm, Caprini:2019egz}. The detectability of these GWs becomes a focal point, as they hold the potential to be observed in various gravitational wave experiments. Not only the production of stochastic GW is a pivotal point for a first order EWPT but also the high production/release of entropy into the plasma which can dilute any preexisting frozen out species like dark matter, baryon asymmetry, etc. is a matter of severe importance. This release of entropy is almost negligible within the framework of the SM because EWPT is a of crossover in nature, \cite{Chaudhuri:2017icn}. But once the SM is extended, even by a singlet, there is a significant production of entropy during EWPT, \cite{Chaudhuri:2022sis, Chaudhuri:2021ibc}.

This pursuit of gravitational wave detection has been a longstanding practice, especially in the context of exploring diverse beyond the Standard Model (BSM) frameworks \cite{Kikuta:2014eja,Costa:2022oaa, Ahriche:2018rao, Shajiee:2018jdq, Alves:2018jsw, Chatterjee:2022pxf, Wang:2022lxn, Kakizaki:2015wua, Hashino:2016rvx, Hashino:2016xoj, Hashino:2018zsi}. Notably, it serves as a complementary avenue alongside traditional collider searches \cite{Ramsey-Musolf:2019lsf, Friedrich:2022cak, Hashino:2018wee}, offering a multifaceted approach to unravel the mysteries of theoretical frameworks beyond the Standard Model.

In this work we consider the next-to-minimal two Higgs doublet model (N2HDM), which is a singlet extension to the two Higgs doublet model (2HDM), as the extension to the SM. By scanning the parameter space for the first order phase transition (FOPT) at the electroweak scale we study the production of the stochastic GW and predict their possible detection in the future GW observations. 

This paper is organised as follows. In Section \ref{sec:model}, we explain the scalar sector of the N2HDM potential. Then, in Section \ref{eff:pot}, we discuss the thermal and one-loop corrections to the scalar potential. Section \ref{GW} is divided into two subsections. Subsection \ref{sec:entropy} deals with the entropy production scenarios due to EWPT within the framework of the N2HDM and subsection \ref{GW section} focuses on the study of the stochastic gravitational waves production due to EWPT. This section is followed by the conclusion \ref{conc}, we summarize our analysis.

\section{Model}
\label{sec:model}
Extensions to the Higgs sector, such as the two-Higgs-doublet model (2HDM) or the Next-to-Minimal-2HDM (N2HDM), offer a diverse collider phenomenology, deeply intertwined with early Universe physics \cite{Branco:2011iw, Grzadkowski:2009iz, Chen:2013jvg, Drozd:2014yla, Muhlleitner:2016mzt, Dorsch:2013wja, Dorsch:2014qja, Basler:2016obg, Dorsch:2017nza, Bernon:2017jgv, Heinemeyer:2021msz, Biekotter:2019kde}. In this study, we specifically investigate this connection within the framework of the N2HDM, with a focus on type II. Our analysis delves into the thermal history, elucidating the evolution of Higgs fields in the early Universe. We reveal that across significant portions of the N2HDM parameter space, the thermal history diverges notably from the conventional scenario of electroweak symmetry breaking occurring at an early Universe temperature $T$ of around $\mathcal{O}(100)~\rm{GeV}$. Two critical phenomena, namely electroweak symmetry non-restoration and vacuum trapping, emerge as pivotal factors shaping this distinct thermal evolution.

The N2HDM tree-level scalar potential consists of two $\text{SU(2)}_{L}$ Higgs doublets
$\Phi_{1}$ and $\Phi_{2}$ and the real singlet field $\Phi_{S}$. A $\mathbb{Z}_{2}$ symmetry is imposed, which is explicitly broken and transforms the fields as $\Phi_{1}\rightarrow\Phi_{1},\,\Phi_{2}\rightarrow-\Phi_{2},\,\Phi_{S}\rightarrow\Phi_{S}$. At this moment we do not impose the extra $\mathbb{Z}'_{2}$ symmetry as mentioned in \cite{Muhlleitner:2016mzt, Biekotter:2021ysx}. This allows us to retain the cubic terms involving $\Phi_{S}$ in the scalar potential, which takes the form:
\begin{align} \label{Tree-Level Potential}
	V_{\text{tree}}&=m_{11}^{2}\left|\Phi_{1}\right|^{2}+m_{22}^{2}\left|\Phi_{2}\right|^{2}-m_{12}^{2}\left(\Phi_{1}^{\dagger}\Phi_{2}+\text{h.c.}\right)+\frac{\lambda_{1}}{2}\left(\Phi_{1}^{\dagger}\Phi_{1}\right)^{2}+\frac{\lambda_{2}}{2}\left(\Phi_{2}^{\dagger}\Phi_{2}\right)^{2} \notag \\
	&+\lambda_{3}\left(\Phi_{1}^{\dagger}\Phi_{1}\right)\left(\Phi_{2}^{\dagger}\Phi_{2}\right)+\lambda_{4}\left(\Phi_{1}^{\dagger}\Phi_{2}\right)\left(\Phi_{2}^{\dagger}\Phi_{1}\right)+\frac{\lambda_{5}}{2}\left[\left(\Phi_{1}^{\dagger}\Phi_{2}\right)^{2}+
	\mathrm{h.c.}\right] \notag \\
	&+\frac{1}{2}m_{S}^{2}\Phi_{S}^{2}+a_1^3\Phi_{S}-a_2\Phi_{S}^3+a_3\left(\Phi_{1}^{\dagger}\Phi_{1}\right)\Phi_{S}  \notag \\
 &+\frac{\lambda_{6}}{8}\Phi_{S}^{4}+\frac{\lambda_{7}}{2}\left(\Phi_{1}^{\dagger}\Phi_{1}\right)\Phi_{S}^{2}+a_4\left(\Phi_{2}^{\dagger}\Phi_{2}\right)\Phi_{S}+\frac{\lambda_{8}}{2}\left(\Phi_{2}^{\dagger}\Phi_{2}\right)\Phi_{S}^{2}.
\end{align}
It is clear from ~\eqref{Tree-Level Potential} that $a_i~(i=1,2,3,4)$ has the dimension of mass/temperature. 
The $\mathbb{Z}_{2}$ symmetry in ~\eqref{Tree-Level
	Potential}, serves to safeguard against flavor-changing neutral currents (FCNCs) at the lowest order when extended to the Yukawa sector. However, this symmetry is softly broken by the presence of the $m_{12}^{2}$ term. 
In the third and fourth line of the tree-level potential \eqref{Tree-Level Potential}, we incorporate the impact of the singlet field. We explore the scenario where $\Phi_{S}$ obtains a vacuum expectation value (vev). To analyze this further, we expand the fields around the EW minimum as follows:
\begin{equation}
	\Phi_{1}=\left(\begin{array}{c}
		\phi_1^+ \\
		\frac{1}{\sqrt{2}}\left(v_{1}+\rho_{1}+i\eta_{1}\right)
	\end{array}\right),\quad\Phi_{2}=\left(\begin{array}{c}
		\phi_2^+ \\
		\frac{1}{\sqrt{2}}\left(v_{2}+\rho_{2}+i\eta_{2}\right)
	\end{array}\right),\quad\Phi_{S}=v_{S}+\rho_{3},
	\label{Higgs fields expansions}
\end{equation}
where $v_1$, $v_2$ and $v_S$ are the field vevs for the Higgs doublets and the singlet field, respectively, at zero temperature. The fields $\phi_i^+$ (where $i=1,2$) represent the complex scalar components linked to the two Higgs doublets. The CP-even and odd fields are denoted by $\rho_I$ (with $I=1,2,S$) and $\eta_i$, respectively. The vevs of the doublets are represented by $v_{1}$ and $v_{2}$, defining the electroweak (EW) scale $v=\sqrt{v_{1}^{2}+v_{2}^{2}}$, which is approximately $246$ GeV. 
%The symmetry breaking involving $\Phi_S$ can be explained by the following figure:
%\begin{figure}[h!]
%	\centering
%	\includegraphics[scale=0.7]{S3GW.pdf}
%	\caption{An example of a strong FOPT involving the $\Phi_S$ which arises due to the presence of the cubic term in the potential in \eqref{Tree-Level Potential}.}
%	\label{FOPT}
%\end{figure}

With the introduction of another discrete $\mathbb{Z}'_{2}$ symmetry, the $a_i$ vanishes and is the best “coordinate
frame” for the singlet as it makes explicitly apparent such symmetry. In case the singlet does not acquire a vev, then it can be a suitable candidate for a dark matter particle but we do not divulge into the study of dark matter. 

At this point, it is important to lay out some differences between the 2HDM and the N2HDM. Not only does the N2HDM has more degrees of freedom compared to the 2HDM because of the extra singlet scalar but it exhibits a different vacuum structure as well. At tree level, the general vacuum structure of the 2HDM allows for three different possible vacua that are given by the normal EW-breaking vacuum, a CP-breaking and a charge-breaking (CB) vacuum \cite{Ivanov:2007de, Barroso:2007rr}. Given that this statement might not remain valid at higher orders or could be disrupted by finite temperature effects, we entertain the possibility of a more versatile vacuum structure, a characteristic inherent to the N2HDM.

N2HDM exhibits a different vacuum structure compared to 2HDM (for a detailed review of the vacuum structure of the N2HDM, please refer to \cite{Basler:2019iuu}). General properties of vacuum structure has been studied in \cite{Ferreira:2019iqb}). As the loop corrections and finite temperature effects may change the vacuum configuration, the doublets and singlet can be expressed in the general vacuum structure as:
\begin{equation}
	\Phi_{1}=\frac{1}{\sqrt{2}}\left(\begin{array}{c}
		\rho_1+i\eta_1 \\
		\left(\zeta_{1}+\omega_{1}+i\psi_{1}\right)
	\end{array}\right),\quad\Phi_{2}=\frac{1}{\sqrt{2}}\left(\begin{array}{c}
		\rho_2+\omega_{CB}+i\eta_2 \\
		\left(\zeta_{2}+\omega_{2}+i(\psi_{2}+\omega_{CP})\right)
	\end{array}\right),\quad\Phi_{S}=\zeta_{3}+\omega_{S},
	\label{vac struct expansion}
\end{equation}
The Higgs fields are expanded in terms of the charged sectors $\rho_i$ and $\eta_i$ ($i=1,2$) and the neutral CP-even and CP-odd fields $\zeta_i$ and $\psi_i$ respectively. Our objective is to secure a minimum that conserves both electric charge and CP while yielding three CP-even massive scalars.  This requirement, from the standpoint of vacuum structure suggests $\omega_1\neq0,~\omega_2\neq0~ \rm{and}~\omega_s\neq0$ while the CB and the CP components are zero. This in turn, provides us \eqref{Higgs fields expansions}. Because of the uniqueness of the vacuum structure, the tree level potential itself contributes to the strength of the electroweak phase transition by enhancing it. This, in turn, increases the energy density of the stochastic GWs produced due to this phenomenon as we shall see in the later part of this paper.

%The minimization (or tadpole) equations for $v_1$, $v_2$ and $v_S$ read
%\begin{align}
%	\frac{v_{2}}{v_{1}}m_{12}^{2}-m_{11}^{2}=\frac{1}{2}\left(v_{1}^{2}\lambda_{1}+v_{2}^{2}\lambda_{345}+v_{S}^{2}\lambda_{7}\right)
%	\ , \label{min_con_1} \\
%	\frac{v_{1}}{v_{2}}m_{12}^{2}-m_{22}^{2}=\frac{1}{2}\left(v_{1}^{2}\lambda_{345}+v_{2}^{2}\lambda_{2}+v_{S}^{2}\lambda_{8}\right)
%	\ , \label{min_con_2} \\
%	-m_{S}^{2}=\frac{1}{2}\left(v_{1}^{2}\lambda_{7}+v_{2}^{2}\lambda_{8}+v_{S}^{2}\lambda_{6}\right),\label{min_con_3}
%\end{align}
%with $\lambda_{345}\equiv\lambda_{3}+\lambda_{4}+\lambda_{5}.$ 

\section{The One Loop Finite Temperature Effective potential}
\label{eff:pot}
The effective potential $V_{\mathrm{eff}}$, encapsulates the influence of radiative corrections on the scalar potential within the theory, particularly at zero temperature \cite{Quiros:1994dr}. At the one-loop order, $V_{\mathrm{eff}}$ can be expressed as the sum of the N2HDM tree-level potential $V_{\mathrm{tree}}$  \eqref{Tree-Level Potential} and the Coleman-Weinberg potential $V_{\rm CW}$. The Coleman-Weinberg potential is well-established and is defined by:
\begin{equation}
	V_{\text{CW}}(\phi_i) = \sum_{j}\frac{n_{j}}{64\pi^{2}}(-1)^{2s_{i}} \,
	m_{j}(\phi_i)^{4}
	\left[\ln\left(\frac{|m_{j}(\phi_i)^{2}|}{\mu^{2}}\right)-c_{j}\right].
	\label{CW_potential}
\end{equation}
Here, $m_{j}(\phi_i)$ represents the field-dependent tree-level mass of the particle species $j$ in our model, where $n_{j}$ denotes its corresponding number of degrees of freedom and $s_{j}$ represents the particle spin. To compute the zero-temperature loop-corrected scalar masses, we incorporate counter terms into $V_{\mathrm{eff}}$, yielding:
\begin{equation}
V_{\rm CT}=\sum_{i}\frac{\partial V_{0}}{\partial p_{i}}
\delta p_{i}+\sum_{k}(\phi_{k}+v_{k})\delta T_{k} \ ,
\label{eq:introparacounters}
\end{equation}
where $p_{i}$ represents the parameters of the tree-level potential.

%%%%%%%%%%%%%%%%%%%%%%%%%%%%%%%%%%%%%%%%%%%%%%%%%%%%%%%%%%%%%%%%%%%%%%%%%%%%%%%%%%%%%%%%%%%%
\subsection{Finite Temperature Effective Potential}
To explore the thermal evolution of the N2HDM, it's essential to compute the effective potential, accounting for finite temperature corrections. The one-loop effective potential, now written as $V_{\rm tot}(\Phi_1,\Phi_2, \Phi_S,T)$, under finite temperature corrections is expressed as, \cite{Delaunay:2007wb}:
\begin{equation}
	V_{\rm tot}(\Phi_1,\Phi_2, \Phi_S,T)\equiv V_{\rm tree} + V_{\rm CW} + V_{T} \, ,
	\label{1loopeffectivepotential}
\end{equation}
where $V_{T}$ is the one-loop thermal potential,
\begin{equation}
	V_{T} (\phi_i)=\sum_{j} \, \frac{n_{j} \, T^{4}}{2\pi^{2}}\,
	J_{\pm}\left(\frac{m_{j}^2(\phi_i)}{T^{2}}\right)  ,
	\label{thermal_potential}
\end{equation}
with the thermal integrals for fermionic ($J_{+}$) and bosonic ($J_{-}$)
sectors: 
\begin{equation}
	J_{\pm}\left(\frac{m_{j}^2(\phi_i)}{T^{2}}\right) = \mp
	\int_{0}^{\infty}dx\,x^{2}\,\log\left[1\pm\exp\left(-\sqrt{x^{2}+\frac{
			m_{j}^2(\phi_i)  }{T^{2}}}\right)\right]\, ,
	\label{thermalfunctions}
\end{equation}
which vanish as $T\rightarrow 0$ (assuming $m^2_j$ is positive).
In the high temperature limit, it is convenient to expand the thermal functions $J_{\pm}$:
\begin{eqnarray}
	J_{-}(y) &\approx&
	-\frac{\pi^{4}}{45}+\frac{\pi^2}{12}y-\frac{\pi}{6}y^{\frac{3}{2}}-\frac{1}{32}\,
	y^2 \, \text{log}\left(\frac{|y|}{a_b}\right) + \mathcal{O}(y^3)\, ,  \nonumber \\
	J_{+}(y) &\approx&  -\frac{7\pi^{4}}{360}+\frac{\pi^2}{24}y+\frac{1}{32}\, y^2
	\, \text{log}\left(\frac{|y|}{a_f}\right) + \mathcal{O}(y^3) \,  \quad \quad
	\mathrm{for} \, \left| y \right| \ll 1 \, ,
	\label{thermal_expansion}
\end{eqnarray}
with $a_b=\pi^2\text{exp}(3/2-2\gamma_{E})$ and
$a_f=16\pi^2\text{exp}(3/2-2\gamma_{E})$, $\gamma_{E} =  0.57721\ldots$ being the
Euler-Mascheroni constant. In our analysis, we focus solely on the contribution from the top quark, as the impact of other fermions within the SM is relatively minor due to their small Yukawa couplings. Additionally, when the field-dependent masses at a given temperature are significantly smaller than the temperature of the plasma, i.e., $\frac{m_i^2}{T^2}<< 1$, the thermal function allows for a high-temperature expansion. This feature proves highly beneficial for practical applications.

In another scenario, when considering a given plasma temperature, if the field-dependent mass of a particle surpasses the temperature significantly, i.e., $\frac{m_i(\Phi_i)^2}{T^2}>>1$, both bosonic and fermionic thermal functions exhibit behavior akin to exponentially decreasing functions \cite{Delaunay:2007wb}. Consequently, in the thermal effective potential, the contribution of a particle with a field-dependent mass greater than the temperature becomes nearly negligible. The initial terms on the right-hand side of equations \eqref{thermal_expansion} are devoid of classical background fields, rendering them inconsequential in computing the critical temperature, $T_c$.

%%%%%%%%%%%%%%%%%%%%%%%%%%%%%%%%%%%%%%%%%%%%%%%%%%%%%%%%%%%%
\section{Stochastic Gravitational Waves}
\label{GW}
\subsection{First Order Electroweak Phase Transition and Entropy Production} \label{sec:entropy}
The study of electroweak phase transition (EWPT) is important in the study of electroweak baryogenesis (EWBG) and GWs. A model which undergoes a strong first order EWPT can successfully create EWBG and produce stochastic GW during the PT. A strong FOPT is necessary to suppress processes such as $SU(2)_L$ sphalerons, which could potentially erase the baryon asymmetry after its production. A strong first order EWPT can result in the production and release of entropy into the primeval plasma, which dilutes any preexisting baryon asymmetry and the energy density of the GWs. Hence, at this point it is necessary to look into the entropy production due to EWPT in the framework of N2HDM.

It is an established fact that the entropy density within the primordial plasma remains constant throughout cosmological expansion, provided the plasma remains in a state of thermal equilibrium with negligible chemical potentials for all particle species. It can be expressed as:
\begin{equation} \label{entropy}
    s=\frac{\mathcal{P}+\rho}{T}a^3=\rm{const,}
\end{equation}
where $\mathcal{P}~\rm{and} \rho$ are the pressure and the energy density and $T$ is the temperature of the plasma while $a(t)$ is the cosmological scale factor. 

The Lagrangian density for the electroweak theory in N2HDM can be expressed as:
\begin{equation} 
\mathcal{L}= \mathcal{L}_f  + \mathcal{L}_{\rm Yuk} + \mathcal{L}_{\rm gauge, kin}  + \mathcal{L}_{\rm Scalar}. 
\label{Eq: Total lagrangian}
\end{equation}
In \eqref{Eq: Total lagrangian}, $\mathcal{L}_f$ is the kinetic term which originates from the fermionic sector of the model and is given by:
\begin{eqnarray}
\mathcal{L}_f &=&
 \sum_{j}i (\bar{\Psi}^{(j)}_L \slashed{D}\Psi^{(j)}_L + \bar{\Psi}^{(j)}_R \slashed{D}\Psi^{(j)}_R  ) \notag \\ &=&i\bar{\Psi}_L \gamma^\xi (\partial_\mu + ig W_\mu + i g'Y_L B_\mu) \Psi_L   \nonumber \\&& + i\bar{\Psi}_R  \gamma^\mu (\partial_\mu + ig W_\mu + i g'Y_R B_\mu) \Psi_R.     \label{Eq: Fermionic Lagrangian}
 %\mathcal{L}_f &=&
 %\sum_{j}i (\bar{\Psi}^{(j)}_L \slashed{D}\Psi^{(j)}_L + \bar{\Psi}^{(j)}_R \slashed{D}\Psi^{(j)}_R  ) \\ &=&i\bar{\Psi}_L \gamma^\xi (\partial_\xi + ig W_\xi + i g'Y_L B_\xi) \Psi_L   \nonumber \\&& + i\bar{\Psi}_R  \gamma^\xi (\partial_\xi + ig W_\xi + i g'Y_R B_\xi) \Psi_R,     \label{Eq: Fermionic Lagrangian}
\end{eqnarray}
Here $L$ and $R$ denote the left and right chiral fields of the fermion, respectively. $\slashed{D}$ represents the gauge covariant derivative in Feynman notation, and $j$ encompasses all fermionic species (i.e., the field $\Psi_j$). The symbol $g$ represents the coupling constant. The partial derivative goes as $\partial _0 = d/dt$ and $\mu$ runs from $0$ to $3$, \cite{Karmakar:2019vnq}. The second term on the R.H.S of \eqref{Eq: Total lagrangian} originates from the Yukawa interaction and is given by:
\be
\mathcal{L}_{\rm Yuk}= -\left[ y_e \bar{e_R} \Phi_a^\dagger L_L + y_e^* \bar{L_L} \Phi_a^\dagger e_R  + \cdots \right],
\ee
where $y_e$ is dimensionless and  complex constant. $\Phi_a$ (where $a=1,2$) represents an $SU(2)_L$ doublet, and to maintain gauge invariance, it's coupled with another $SU(2)_L$ fermion, $L_L$. Similarly, $e_R$ and other fermions such as quarks and neutrinos are right chiral fields, consistent with the aforementioned coupling. $\mathcal{L}_{\rm gauge, \,kin} $ represents $U(1)$ invariant  kinetic term of four gauge bosons ($W^i, \, i=1,2,3$, and $B$). It can be written as:
\begin{eqnarray}
\mathcal{L}_{\rm gauge,\, kin} = -\frac{1}{4}G^i_{\mu \nu}{G^i}^{\mu \nu}-\frac{1}{4}F^B_{\mu \nu}{F^B}^{\mu \nu},          \label{Gauge-kinetic Lagrangian}
\end{eqnarray}
where $G^i_{\mu \nu}=\partial_\mu W^i_\nu-\partial_\nu W^i_\mu - g \epsilon^{ijk}W_\mu^j W_\nu^k$ and $F^B_{\mu \nu}=\partial_\mu B_\nu-\partial_\nu B_\mu$ 
with $\epsilon_{ijk}$ is the Levi-Civita symbol. The last term in \eqref{Eq: Total lagrangian} deals with the scalar sector which includes the two Higgs doublet and the scalar singlet particle. It can be written as:
\begingroup\makeatletter\check@mathfonts
\def\maketag@@@#1{\hbox{\m@th\normalsize\normalfont#1}}%
\begin{eqnarray}
\mathcal{L}_{\rm Scalar} &=& (D^\mu \Phi_1)^\dagger (D_\mu \Phi_2) + (D^\mu \Phi_1)^\dagger (D_\mu \Phi_2)+ 2((D^\mu \Phi_1)^\dagger (D_\mu \Phi_S)+(D^\mu \Phi_2)^\dagger (D_\mu \Phi_S))   \nonumber \\  
&& -V_{\rm tot}(\Phi_1,\Phi_2, \Phi_S,T),
\label{Eq:eq1}
\end{eqnarray} 
\endgroup
where 
\begin{equation}
    V_{\rm tot}(\Phi_1,\Phi_2, \Phi_S,T)\equiv V_{\rm tree} + V_{\rm CW} + V_{T},
\end{equation}
where all the individual components of the RHS are mentioned in \eqref{Tree-Level Potential}, \eqref{CW_potential} and \eqref{thermal_potential}.

\begin{figure}[h!]
	\centering
	\includegraphics[scale=0.8]{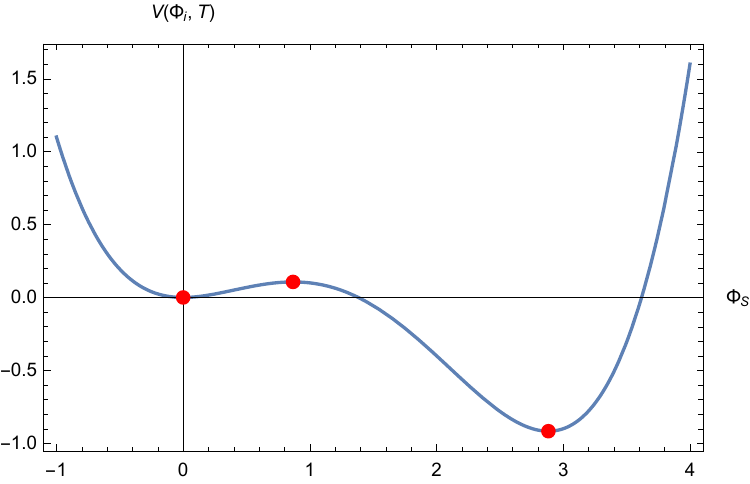}
	\caption{Strong EWPT can be seen in this case when the symmetry breaking takes place. The plot is made with respect to $\Phi_S$ as it contains the cubic term in the potential in \eqref{Tree-Level Potential}. The parameter space of BP1 \eqref{tab:Entropy_BP} has been followed here.}
	\label{fig:evol1}
\end{figure}

As the temperature of the universe decreases, a secondary local minimum emerges. When the plasma temperature reaches $T_c$, this secondary local minimum adopts the form $(\left<\Phi_1\right>=v_1,\left<\Phi_2\right>=v_2,\left<\Phi_S\right>=v_S)$ and becomes degenerate with the global minimum at $(\left<\Phi_1\right>=\left<\Phi_2\right>=\left<\Phi_S\right>=0)$. The structure of the potential after the symmetry breaking is shown in figure \ref{fig:evol1}. The critical temperature is determined by the following expression:
 \begin{equation}
     V_{\rm tot}\left(\Phi_1=0,\Phi_2=0,\Phi_S=0,T_c \right)= V_{\rm tot}\left(\Phi_1=v_1,\Phi_2=v_2,,\Phi_S=v_S, T_c \right).
 \end{equation}

Assuming the field to be homogeneous, the energy density can be written as:
\begin{align}
	\rho =&\partial^0 \Phi_a^\dagger \partial^0\Phi_a - (\mathcal{W}^0 \Phi_a)^\dagger \mathcal{W}^0\Phi_a 
	- (\mathcal{W}^j \Phi_a)^\dagger \mathcal{W}_j\Phi_a  \nonumber \\
	&+\lt[V_{\rm tot}(\Phi_1,\Phi_2,\Phi_S, T)  - \mathcal{L}_{\rm gauge, kin} - \mathcal{L}_f  - \mathcal{L}_{\rm Yuk}  ,
	\rt]     \label{Eq: Expression for energy desnity}
\end{align}
where  $\mathcal{W}_\mu=gT^iW^i_\mu + g'Y B_\mu$. Because of the homogeneity and isotropy condition the spatial derivatives of the Higgs fields vanishes. Following \eqref{entropy}, the sum of pressure density and energy density of the plasma is given by:
\begin{eqnarray}
\rho + \mathcal{P}= 2\partial^0\Phi_a \partial^0 \Phi_a^\dagger - i ({\mathcal{W}^0}\Phi_a)^\dagger (\partial_0 \Phi_a) + i (\partial^0{\Phi_a}^\dagger)\mathcal{W}_0\Phi_a.
%\\
%+  {\color{purple} \mathcal{L}_{\rm Yuk} } \right]   
\label{Faltu equation 5}
\end{eqnarray}
Near the electroweak phase transition, $g_*\sim 110$ represents the effective number of relativistic particles. This value varies with temperature and diminishes as the cosmological cooling proceeds. \eqref{Eq: Expression for energy desnity} remains valid in the scenario of instantaneous thermalization. Hence it can be rewritten as:
\be \label{eq: modified energy equation}
\rho =\Dot{\Phi}_{a, {\rm min}}^2 
+  V_{\rm tot}(\Phi_1,\Phi_2, \Phi_S T) + \frac{g_* \pi^2}{30} T^4.
\ee
The final term on the right-hand side of \eqref{eq: modified energy equation} accounts for the energy density of relativistic particles that have yet to acquire mass by the time of the electroweak phase transition (EWPT). This includes contributions from Yukawa interactions between fermions and Higgs bosons, as well as the energy density associated with fermions, gauge bosons, and the interaction between Higgs bosons and gauge bosons.

Before the onset of EWPT, $T\gg \lt(\mu_{\rm chm. \, pot.} -m \rt)$, where $\mu_{\rm chm.\, pot.}$ is the chemical potential of the particle species and it is assumed to be negligible for fermions. The entropy conservation law for the relativistic particles is:
\begin{eqnarray}
\label{eq:rho+P}
%\lt( \rho_r +P_r \rt)
s\lt(t \rt)=\frac{2}{45} \pi^2 g_{\star,s} \, T\lt(t \rt)^3,
\end{eqnarray}
where $g_{\star,s}$ is effective degrees of freedom in the entropy which varies with time and temperature. Although it's influenced by the constituents of the primordial hot mixture, the effective number of relativistic degrees of freedom, denoted as $g_\star$, might not always align with the count of relativistic species. This discrepancy is discussed further in \cite{Lunardini:2019zob}. However, for the purposes of this project, we assumed $g_{\star,s}\approx g_\star$. From \eqref{entropy} and \eqref{eq:rho+P}, we derive $T \sim a^{-1}$, indicating that temperature scales inversely with the scale factor. As the universe expands and cools, it reaches a state of thermal disequilibrium at some stage. Consequently, the value of $s$ and, correspondingly, $g_\star \left(T\right) a^3 T^3$ may have increased, as entropy can either rise or remain constant.

As the temperature decreases from $T_c$ during EWPT, specific components of the relativistic plasma reach their decoupling temperatures. At this point, these components transition from being relativistic to non-relativistic and acquire mass. The temperature at which decoupling occurs hinges upon the masses of the components and their respective coupling constants. This decoupling process leads to alterations in the effective number of relativistic degrees of freedom, $g_{\star}$, within the plasma. EWPT presents a potential scenario for Electroweak Baryogenesis to occur, establishing thermal disequilibrium in the universe according to Sakharov's principle \cite{Sakharov:1967dj}.

As a consequence, the law of entropy conservation is violated, resulting in a net increase in entropy. This increase in entropy is primarily driven by the decoupling process, leading to a change in the effective number of relativistic degrees of freedom, $g_\star$, within the relativistic plasma. The predominant contribution to this increase in entropy stems from the presence of the heaviest particle with a mass $m(T)$ below the prevailing temperature $T$. Consequently, within this temperature range, the product $g_\star (T)T$ remains constant, causing the entropy to rise proportionally to $a^3T^3$. Since the final temperature $T=m(T_f)$ (below which a new particle begins to dominate) is independent of $g_{\star}$, we observe an increase in entropy against the backdrop of $g_{\star}a^3 T^3$. 

For brevity, we assume the oscillations of $\phi$ around $\phi_{min}$ undergo rapid damping. Consequently, we set $\dot \phi = \dot \phi_{min}$ and disregard higher-order terms, as the evolution of $\phi_{min}$ is primarily driven by the gradual expansion of the universe, which occurs at a much slower pace.
This simplification results in a single differential equation for the temperature (or scale factor). In principle, this can be numerically computed by solving the relevant Klein-Gordon equation, accounting for damping effects caused by particle production.

To compute the entropy production resulting from the EWPT, which subsequently impacts the amplitude of stochastic gravitational waves generated during the EWPT, it's essential to solve the evolution equation governing the conservation of energy density,
\begin{equation} \label{fried}
\Dot{\rho}=-3\mathcal{H}(\rho+\mathcal{P}).  
\end{equation}

In the subsequent analysis, we utilize the {\tt cosmoTransitions} package, \cite{Wainwright:2011kj}, which specializes in handling a range of properties and functionalities associated with phase transitions. This package is employed to compute essential parameters such as the critical temperature ($T_c$) and the vacuum expectation value ($v_c$), as well as $V_{\rm eff}(T)$, for each of the benchmark points under consideration. In order to achieve a first order phase transition the ratio $v_c/T_c \gg 1$.  The set of benchmark points (BPs) are shown in Table \ref{tab:Entropy_BP}

\begin{figure}[h!]
  \centering
  \begin{minipage}[b]{0.4\textwidth}
    \includegraphics[width=\textwidth]{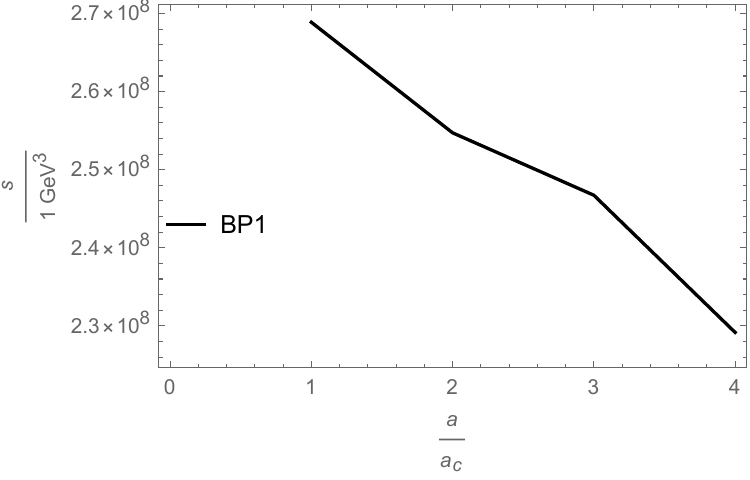}
  %  {figh300b.pdf}    
  \end{minipage}
      \begin{minipage}[b]{0.4\textwidth}
    \includegraphics[width=\textwidth]{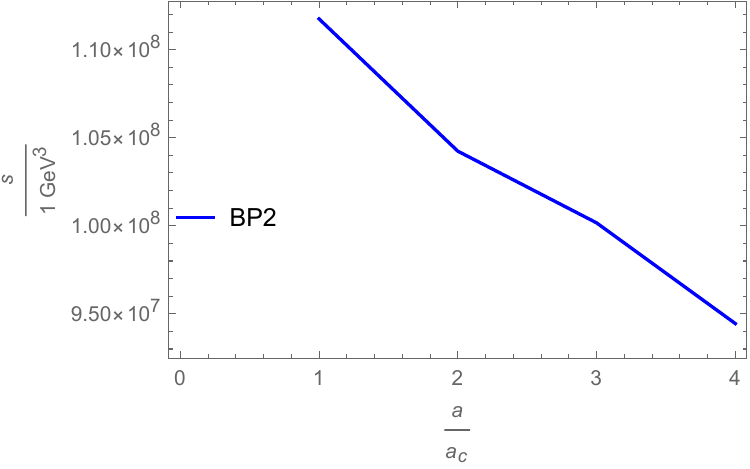}
    %{figh600b.pdf}
      \end{minipage}
      \begin{minipage}[b]{0.6\textwidth}
    \includegraphics[width=\textwidth]{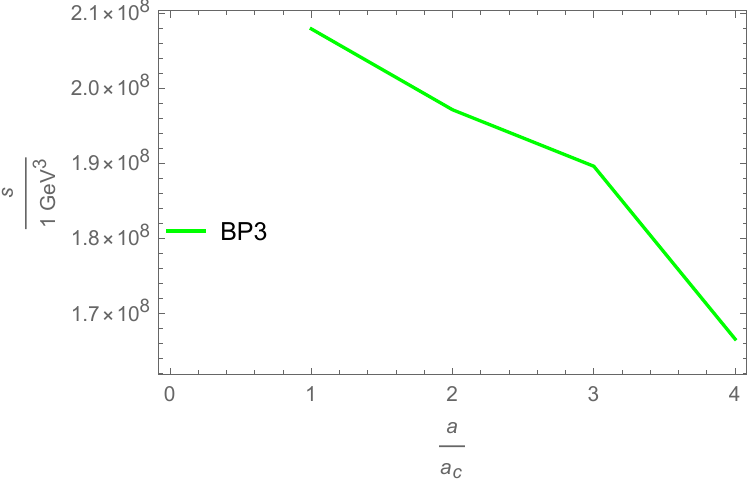}
    %{figh600b.pdf}
      \end{minipage}
      \begin{minipage}[b]{0.4\textwidth}
    \includegraphics[width=\textwidth]{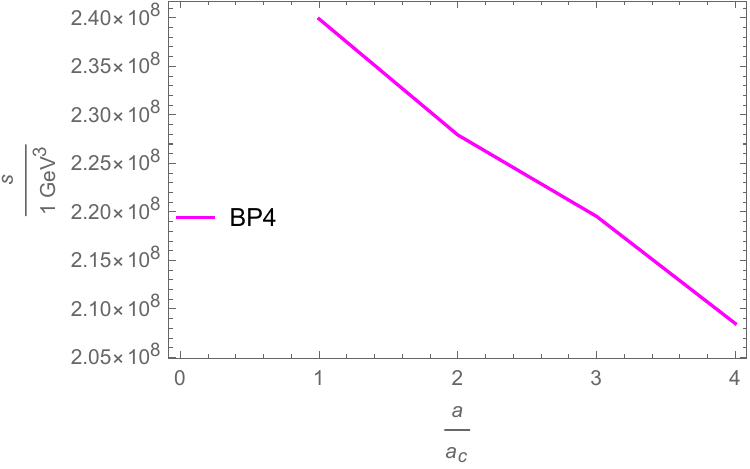}
    %{figh600b.pdf}
      \end{minipage}
      \begin{minipage}[b]{0.4\textwidth}
    \includegraphics[width=\textwidth]{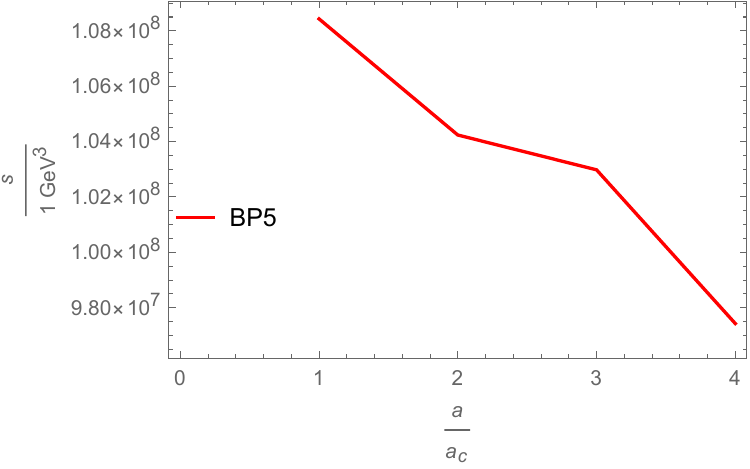}
    %{figh600b.pdf}
      \end{minipage}
  \caption{The figures show the evolution of entropy, which is made dimensionless for brevity as a function of the ratio of scale factor to the scale factor corresponding to the transition temperature for all different BPs presented in Table \ref{tab:Entropy_BP}.}
 \label{entropy_evolution}
 \end{figure}

\begin{table}[h!]
\centering
\begin{tabular}{| c | c | c |c |c |c |c |c |c |c |c |c |c |}
\hline 
~BPs~ & ~$\lambda_1$~ & ~$\lambda_2$~& ~$\lambda_3$~& ~$\lambda_4$ ~& ~$\lambda_5$~& ~$a_1$~& ~$a_2$~& ~$a_3$~& ~$\lambda_6$~& ~$\lambda_7$~& ~$a_4$~& ~$\lambda_8$ \\
\hline
\hline
~BP1~ & ~1.28~ & ~1.28~ & ~0.7~ & ~0.37~ & ~0.37~ & ~0.5~ & ~0.5~ & ~0.1~ & ~0.5~ & ~0.5~ & ~0.1~ & ~0.5 \\
\hline
~BP2~ & ~1.28~& ~0.002~& ~0.7~& ~0.24~& ~0.24~& ~0.7~& ~0.7~& ~0.1~& ~1.5~& ~1.0~& ~0.1~& ~0.5 \\\hline
~BP3~ & ~0.263~& ~0.258~& ~1.06~& ~0.3~& ~1.35~& ~0.1~& ~0.5~& ~0.5~& ~1.35~& ~1.35~& ~0.5~& ~1.35  \\ \hline
~BP4~ & ~1.14~ & ~0.37~& ~0.5~& ~0~& ~0~& ~1.2~& ~1.2~& ~0.8~& ~1.4~& ~1.4~& ~0.8~& ~0.65  \\ \hline
~BP5~ & ~ 4.13 ~ & ~ 0.22~& ~4.15~& ~0~& ~0~& ~1.8~& ~1.8~& ~1.2~& ~1.6~& ~1.6~& ~1.2~& ~1.6  \\ \hline
\end{tabular}
\caption{Estimates of the various parameters in \eqref{Tree-Level Potential}. Please note that the $\lambda_i,~ (i=1-8))$ are dimensionless constants while $a_i,~(i=1-4))$ are in GeV.}
\label{tab:Entropy_BP}
\end{table}
After numerically solving \eqref{fried} for the various BPs in Table \ref{tab:Entropy_BP}, we have plotted out both the evolution of entropy as a function of the scale factor $a/a_c$ and  the entropy release (or the net increase in entropy production) for each BPs have been calculated and plotted in figure \ref{entropy_evolution}. The values of $T_c$ and the entropy release is provided in Table \ref{tab:Tc}.

%\begin{figure}[h!]
%\centering
 % \begin{minipage}[b]{0.4\textwidth}
  %  \includegraphics[width=\textwidth]{BP1.pdf}
  %  {figh300b.pdf}    
  %\end{minipage}
   %   \begin{minipage}[b]{0.4\textwidth}
    %\includegraphics[width=\textwidth]{BP2.pdf}
    %{figh600b.pdf}
     % \end{minipage}
      %\begin{minipage}[b]{0.6\textwidth}
    %\includegraphics[width=\textwidth]{BP3.pdf}
    %{figh600b.pdf}
     % \end{minipage}
      %\begin{minipage}[b]{0.4\textwidth}
    %\includegraphics[width=\textwidth]{BP4.pdf}
    %{figh600b.pdf}
     % \end{minipage}
      %\begin{minipage}[b]{0.4\textwidth}
    %\includegraphics[width=\textwidth]{BP5.pdf}
    %{figh600b.pdf}
     % \end{minipage}
  %\caption{The figures show the entropy production  for all different BPs presented in Table \ref{tab:Entropy_BP}. The exact values of the $T_c$s and the entropy production rate is given in Table \ref{tab:Tc}.}
 %\label{entropy_plot}
 %\end{figure}

\begin{table}[h!]
\centering
\begin{tabular}{| c | c | c |}
\hline 
~BPs~ & ~$T_c(\rm{GeV})$~ & ~$\frac{\delta s}{s}($\%$)$~  \\
\hline
\hline
~BP1~ & ~129~ & ~30~ \\
\hline
~BP2~ & ~96~& ~40~ \\\hline
~BP3~ & ~119~& ~34~  \\ \hline
~BP4~ & ~125~ & ~33~  \\ \hline
~BP5~ & ~ 89 ~ & ~ 44 ~  \\ \hline
\end{tabular}
\caption{Values of the critical/transition temperature $T_c$ and entropy production corresponding to BPs in Table \ref{tab:Entropy_BP}.}
\label{tab:Tc}
\end{table}

We can see from Figure \ref{entropy_evolution} that during the time of the EWPT, there is a significant increase in the entropy  of the plasma. The entropy starts increasing from the onset of the PT and when the PT comes to an end, it becomes a constant value. By constant value, it is meant that the entropy curve follows the regular $~a^{-3}$ nature, with no change in the slope of the curve. In terms of the net entropy released, it can be said that during the phase transition, the entropy density increased by a certain amount, provided in Table \ref{tab:Tc} and when the period of phase transition ended, it became constant. It is evident from Table \ref{tab:Tc}  that stronger the phase transition, the larger is the production of entropy. In the Table \ref{tab:Tc}, it can be seen that the strength of the phase transition is strongest for BP5, which corresponds to $\sim 44\%$ release in entropy compared to BP1, which has the least strength among the $5$ BPs, which corresponds to $\sim 30\%$ increase in the entropy of the plasa after the EWPT. The primary factor behind this surplus in entropy production stems from the additional scalar sector generated in the N2HDM, which notably contribute to this phenomenon. Notably, contributions from lighter particles such as electrons and neutrinos mirror those observed in the SM. In the next section we will show how the strength of the phase transition is related to the production of stochastic GW.

%%%%%%%%%%%%%%%%%%%%%%%%%%%%%%%%%%%%%%%%%%%%%%%%%%%%%%%%%%%%%%%%%%%%%%%%%%%
\subsection{Stochastic Gravitational Waves} \label{GW section}
A first-order phase transition occurring in the early universe serves as a potential source of stochastic GW. Typically, this transition happens roughly $\sim 10^{-11}~\rm{sec}$ after the big bang, preceding the initiation of Big Bang Nucleosynthesis. As the temperature of the universe decreases, it undergoes a first-order phase transition at $T=T_c$. Above this critical temperature, the universe's symmetry is intact. However, as the temperature drops below $T_c$, a second degenerate minimum emerges, signaling the onset of a first-order phase transition. During this transition, latent heat is released, with a portion contributing to the generation of GWs, while the remainder is absorbed by the plasma. Bubble nucleation governs the progression of a FOPT, making the calculation of the nucleation temperature pivotal for determining essential phenomenological parameters crucial for estimating gravitational wave spectra. As nucleation occurs at temperatures below $T_c$, the probability of tunneling $\Gamma (T)$ from the false vacuum to the true one can be expressed as given in \cite{Grojean:2006bp},
\beq\label{eq:tunnelling-prob}
\Gamma (T) \approx T^4 \left( \frac{S_E}{2 \pi T} \right)^{3/2}\; e^{-\frac{S_E}{T}},
\eeq
where $S_E$ is the Euclidean action, otherwise known as the bounce action. It is given by, \cite{Linde:1981zj}:
\be \label{eq:bounce-action}
S_E = \int_{0}^{\infty} 4\pi r^2 dr \left( V_{T} (\phi, T) + \frac{1}{2}\;\left( \frac{d\phi(r)}{dr} \right)^2  \right).
\ee
Here $\phi$ is the scalar dynamical field whose classical solution can be represented in terms of the radial coordinate $r$ as:
\beq\label{eq:phivariationr}
\frac{d^2 \phi}{dr^2} + \frac{2}{r} \frac{d\phi}{dr} = \frac{dV_{T} (\phi, T)}{dr},
\eeq
More details can be found in \cite{Affleck:1980ac,Linde:1977mm}. The necessary boundary conditions to solve \eqref{eq:phivariationr} are:
\begin{equation} \label{bc:1}
    \frac{d\phi}{dr} = 0~\rm{when}~ r \rightarrow 0,
\end{equation}
and
\begin{equation} \label{bc:2}
    \phi(r) \rightarrow \phi_{\rm false}~\rm{when}~ r \rightarrow \infty.
\end{equation}
In \eqref{bc:2} $\phi_{\rm false}$ denotes the four dimensional field values at the false vacuum. The package {\tt cosmoTransitions} has been used to solve \eqref{eq:phivariationr}. The key parameters crucial for estimating gravitational wave spectra originating from first-order phase transitions are the relative change in energy density during the phase transition ($\alpha$) and the inverse of the duration of the phase transition ($\beta$). Both $\alpha$ and $\beta$ are defined at the nucleation temperature $T_n$. The parameter $\alpha$ is calculated as:
\beq\label{eq:alphaP}
\alpha = \frac{\Delta \rho}{\rho_{\rm rad}},
\eeq
where $\rho_{\rm rad}$ is the radiation energy density and $\Delta \rho$ is the released latent heat released during the phase transition, \cite{Kamionkowski:1993fg}. It is given by, \cite{Kehayias:2009tn}:
\beq\label{eq:delrho}
{\Delta \rho = \left[ V_T(\phi_0, T)-T\frac{dV_T(\phi_0, T)}{dT}  \right]_{T = T_n} - \left[ V_T(\phi_n, T)-T\frac{dV_T(\phi_n, T)}{dT}  \right]_{T = T_n}.}
\eeq
$\phi_0$ and $\phi_n$ are the field values at false and true vacuum and $V_T(\phi, T)$ is the finite temperature effective potential and in our case it is same as \eqref{1loopeffectivepotential}. The $\beta$ parameter is defined as:
\beq\label{eq:betaP}
\frac{\beta}{H_{\ast}} = T \frac{d}{dT} \left( \frac{S_E}{T} \right) \Bigg|_{T = T_{\ast}} \equiv \, T \frac{d}{dT} \left( \frac{S_E}{T} \right) \Bigg|_{T = T_n},
\eeq
where $H_{*}$ is the expansion rate of the Universe during the PT and $T_*$ stands for the PT temperature. In our analysis, we have assumed $T_*=T_n$, where $T_n$ is the nucleation temperature. The various values of $\alpha$ and $\beta/H_*$ are shown in Table \ref{tab:GW_obs} for various benchmark points.  
\begin{table}[h!]
\centering
\begin{tabular}{| c | c | c |}
\hline 
~BPs~ & ~$\alpha$~ & ~$\beta/H_{*}$~  \\
\hline
\hline
~BP1~ & ~0.05~ & ~8930.8~ \\
\hline
~BP2~ & ~0.4~& ~1202.5~ \\\hline
~BP3~ & ~0.6~& ~13866.3~  \\ \hline
~BP4~ & ~0.1~ & ~1295.2~  \\ \hline
~BP5~ & ~ 0.75 ~ & ~ 1757.2 ~  \\ \hline
\end{tabular}
\caption{Estimates of the parameters $\alpha$ and $\beta$}
\label{tab:GW_obs}
\end{table}

There are three primany components which contributes to the energy density of the stochastic GW. They are the contributions from the bubble wall collisions, the sound waves and the magneto-hydrodynamic turbulence (MHD). Thus the net energy spectrum of stochastic GW can be approximated as the sum of all the three components:
\beq\label{eq:GWTotal}
\Omega_{\rm GW}h^2 \approx \Omega_{\rm col} h^2 + \Omega_{\rm sw} h^2 + \Omega_{\rm tur} h^2,
\eeq
where the first term on the RHS of \eqref{eq:GWTotal} is the contribution from the bubble wall collisions, the second term corresponds to the sound-waves and the third term gives the contribution from MHD, \cite{Caprini:2015zlo}. Also, $h = H_0/(100 \text{\,km}\cdot \text{sec}^{-1} \cdot \text{Mpc}^{-1})$, where $H_0$ is the value of the Hubble constant at the present day universe, see \cite{DES:2017txv}. The portion of the overall gravitational wave energy density stemming from bubble wall collisions can be determined utilizing the envelope approximation, \cite{Jinno:2016vai}, and can be expressed a function of frequency $f$ as:
\beq\label{eq:GWcoldetails}
\Omega_{\rm col} h^2 = 1.67 \times 10^{-5} {\left(\frac{\beta }{H_{\ast}} \right)^{-2}} \left( \frac{\kappa_c \alpha}{1 + \alpha} \right)^2 \left(  \frac{100}{g^{\ast}} \right)^{1/3} \left( \frac{0.11 v^3_w}{0.42 + v^2_w} \right) \frac{3.8 \left( f/f_{\rm col} \right)^{2.8}}{1 + 2.8 \left( f/f_{\rm col} \right)^{3.8}}\;.
\eeq
Here $v_w$ is the bubble wall velocity and $\kappa_c$ is the efficiency factor of bubble collision. It is given by:
\beq\label{eq:kcfac}
\kappa_c = \frac{0.715 \alpha + \frac{4}{27} \sqrt{\frac{3 \alpha}{2}}}{1 + 0.715 \alpha}.
\eeq
$f_{col}$ is the red-shifted frequency peak and can be written as:
\beq\label{eq:PF1}
f_{\rm col} = 16.5 \times 10^{-6} \left( \frac{f_{\ast}}{\beta} \right) \left( \frac{\beta}{H_{\ast}} \right) \left( \frac{T_n}{100 {\rm \,GeV}} \right) \left( \frac{g^{\ast}}{100} \right)^{1/6} {\rm Hz},
\eeq
where $f_{\ast}/\b$ is called the fitting function and it is given by:
\beq\label{eq:fastbybetadetails}
\frac{f_{\ast}}{\b} = \frac{0.62}{1.8 - 0.1 v_w +v^2_w}.
\eeq
In what follows we will assume $v_w=1$. This is because expanding bubbles can obtain relativistic terminal velocity. The contribution to the energy density of GW from the sound waves is given by, \cite{Hindmarsh:2013xza,Hindmarsh:2016lnk,Hindmarsh:2017gnf}:
\beq\label{eq:GWswpart}
\Omega_{\rm sw} h^2 = 2.65 \times 10^{-6}\; \Upsilon(\tau_{\rm sw}) \left(  \frac{\beta}{H_{\ast}} \right)^{-1} v_w \left( \frac{\kappa_{\rm sw} \alpha}{1 + \alpha} \right)^2 \left( \frac{g^{\ast}}{100} \right)^{1/3} \left( \frac{f}{f_{\rm sw}} \right)^3 \left[ \frac{7}{4 + 3 \left( f/f_{\rm sw} \right)^2} \right]^{7/2},
\eeq
where $\kappa_{\rm sw}$ denotes the efficiency factor for the sound wave contribution, indicating the latent heat transformed into bulk motion of the plasma, thereby emitting gravitational waves. This expression holds true in the limit as the wall velocity approaches $v_w\to 1$, and can be expressed as:
\beq\label{eq:kappasw}
\kappa_{\rm sw} \simeq \left[ \frac{\alpha}{0.73 + 0.083 \sqrt{\alpha} + \alpha} \right].
\eeq
The peak frequency for the contribution of sound waves is denoted by $f_{\rm sw}$ and is given by:
\beq\label{eq:PF2}
f_{\rm sw} = 1.9 \times 10^{-5} \left( \frac{1}{v_w} \right) \left( \frac{\beta}{H_{\ast}} \right) \left( \frac{T_n}{100 {\rm ~GeV}} \right) \left( \frac{g^{\ast}}{100} \right)^{1/6} {\rm Hz}.
\eeq
The suppression of the contribution of the sound waves to the total energy density of the GW energy spectrum is caused by $\Upsilon (\tau_{\rm sw})$, which comes into play due to the finite lifetime of the sound waves and it is given by:
\beq\label{eq:swtimepart}
\Upsilon(\tau_{\rm sw}) = 1 - \frac{1}{\sqrt{1+2 \tau_{\rm sw} H_{\ast}}},
\eeq
with $\tau_{\rm sw}$ being the lifetime of the sound waves. The contribution from the MHD, which arises due to the complete ionization of the plasma, \cite{Caprini:2009yp}, is given by:
\beq\label{eq:GWturpart}
\Omega_{\rm tur} h^2 = 3.35 \times 10^{-4} \left( \frac{\beta}{H_{\ast}} \right)^{-1} v_w \left( \frac{\kappa_{\rm tur} \alpha}{1 + \alpha} \right)^{3/2} \left( \frac{100}{g^{\ast}}\right)^{1/3} \left[ \frac{\left( f/f_{\rm tur} \right)^3}{\left[ 1 + \left( f/f_{\rm tur} \right) \right]^{11/3} \left( 1 + \frac{8 \pi f}{h_{\ast}} \right)} \right],
\eeq
where $h_{\ast} = 16.5 \times \left( \frac{T_n}{100 {\rm ~GeV}} \right) \left( \frac{g^{\ast}}{100} \right)^{1/6} {\rm Hz}$, the inverse Hubble time during GW production, red-shifted to today. The peak frequency $f_{\rm tur}$ is given by,
\beq\label{eq:PF3}
f_{\rm tur} = 2.7 \times 10^{-5} \frac{1}{v_w} \left( \frac{\beta}{H_{\ast}} \right) \left( \frac{T_n}{100 {\rm~GeV}} \right) \left( \frac{g^{\ast}}{100} \right)^{1/6} {\rm Hz.}
\eeq
$\kappa_{\rm tur}$ is defined as $\kappa_{\rm tur}=\epsilon \kappa_{\rm{ sw}}$, where $\epsilon$ stands for the fraction of the bulk motion which is turbulent and previous results, \cite{Caprini:2009yp}, suggests $\kappa_{\rm tur}=0.1 \kappa_{\rm sw}$, which is used in the calculations.

\begin{figure}[h!]
	\centering
	\includegraphics[scale=0.6]{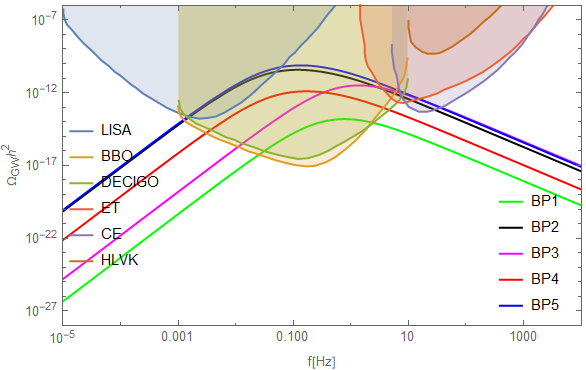}
	\caption{GW spectrum for various benchmark points corresponding to Table \ref{tab:GW_obs}. The results of our calculations are compared with the available sensitivity plots provided by the various GW interferometers for future observations. \cite{Kudoh:2005as,Yagi:2011wg,Nakayama:2009ce,NANOGrav:2023hvm,LISA:2017pwj,LIGOScientific:2014pky,VIRGO:2014yos,KAGRA:2018plz,Crowder:2005nr, Kawamura:2011zz,Punturo:2010zz, Reitze:2019iox}}
	\label{fig:evol}
\end{figure}

Figure \ref{fig:evol} shows the plots of the total energy density of the produced stochastic GWs (equivalent to \eqref{eq:GWTotal}), considering all the aforementioned points. The produced results has been compared with the predictions from various GW interferometers LISA, BBO, DECIGO and others \cite{Kudoh:2005as,Yagi:2011wg,Nakayama:2009ce,NANOGrav:2023hvm,LISA:2017pwj,LIGOScientific:2014pky,VIRGO:2014yos,KAGRA:2018plz,Crowder:2005nr, Kawamura:2011zz,Punturo:2010zz, Reitze:2019iox}.

The parameter $\alpha$ is directly related to the energy released during the phase transition. Therefore, a stronger phase transition should result in a higher value of $\alpha$. It is obvious from the plot in Figure \ref{fig:evol} that the strength of the phase transition corresponding to BP5 is the strongest and hence we see that the amplitude of the GW spectrum has the highest value when BP5 is considered. Similarly, for BP1, we find the strength of the phase transition is the weakest because $\alpha$ has the lowest value corresponding to BP1 and hence this is reflected on the amplitude of the GW spectrum.

\section{Conclusion} \label{conc}
In this work, we studied two implications of a strong first order EWPT within the framework of the next-to-minimal Two Higgs Doublet Model. We considered only one $\mathbb{Z}_{2}$ symmetry imposed on the Higgs doublets and the singlet, thus allowing the singlet to have cubic terms in the potential which gives rise to a strong FOPT inherently, as can be seen in Figure \ref{fig:evol1}. We proceeded to calculate the entropy production due to this EWPT and we found out that there is a significant increase/influx of entropy into plasma during the phase transition. Furthermore, we have studied the production of stochastic gravitational waves which can be produced during the EWPT and we indeed found prominent signatures of GW signals and they overlap well with the sensitivity curves provided by the GW interferometers and are open to detection in the near future.

We saw how the strength of the phase transition effects the entropy production and the production of GW. A stronger PT results in a higher production of both entropy and GW. However, in our calculations, we have assumed the bubble wall velocity $v_w\sim 1$. But a precise calculation and determination of this bubble wall velocity, even though, non-trivial, is possible as shown in \cite{Dine:1992wr,Ignatius:1993qn,Moore:1995si}. This might have an effect on the net production of GW but analysis of \eqref{eq:GWTotal} considering non-unitary values of $v_w$ is beyond the scope of this paper.

\section{Acknowledgement}
The work of AC was supported by the Japan Society for the Promotion of Science (JSPS) as a part of the JSPS Postdoctoral Program (Standard), grant number: 23KF0289. The work of KK was supported by the KAKENHI grant No. 24H01825. AC also thanks Kousik Loho for useful discussion.

\end{document}